\documentclass[twocolumn,aps,prl,amsmath,amssymb,showpacs,preprintnumbers]{revtex4}

\newcommand{\nc}{\newcommand}

\nc{\be}{\begin{equation}}

\nc{\ee}{\end{equation}}

\nc{\bea}{\begin{eqnarray}}

\nc{\eea}{\end{eqnarray}}

\nc{\xx}{\nonumber\\}

\nc{\ct}{\cite}

\nc{\la}{\label}

\nc{\eq}[1]{(\ref{#1})}

\usepackage{color}

\def\ajou#1&#2(#3){\ \sl#1\bf#2\rm(19#3)}

\def\vare{\varepsilon}
\def\zbar{\bar{z}}

\def\half{{\frac{1}{2}}}

\def\zbar{{\bar z}}
\def\[{\left [}
\def\]{\right ]}

\begin{document}
\preprint{HU-EP-06/20}
\preprint{hep-th/0608013}
\title{Instantons and Emergent Geometry}
\author{Hyun Seok Yang}
\email{hsyang@kias.re.kr}
\affiliation{Institut f\"ur Physik,
Humboldt Universit\"at zu Berlin,
Newtonstra\ss e 15, D-12489 Berlin, Germany}

\date{\today}

\begin{abstract}
We show that self-dual electromagnetism in noncommutative spacetime is
equivalent to self-dual Einstein gravity.
\end{abstract}


\maketitle

\section{1. Introduction}

Gravity is a mysterious force; we still don't know what gravity is.
But recent developments from string and M theories gradually reveal a
remarkable and radical new picture about gravity (for recent reviews,
see, for example, \cite{review,mine}):
{\it Gravity may be a collective phenomenon emergent from gauge fields. That is,
  the spin-two graviton might arise as a composite of two spin-one gauge
  bosons.}

Although the emergent gravity is a fascinating approach suggesting
that gravity itself may not be ``fundamental physics'', it is not
easy to realize it from ordinary quantum field theories due to a general no-go
theorem known as the Weinberg-Witten theorem \cite{ww-theorem},
stating that {\it an interacting graviton cannot emerge from an ordinary
quantum field theory in the same spacetime}. One has to notice, however, that
Weinberg and Witten introduced two basic assumptions to prove this
theorem. The first hidden assumption is that gravitons and gauge
fields live in the same spacetime. The second assumption is the
existence of a Lorentz-covariant stress-energy tensor. Thus, to
realize the emergent gravity, one has to relax one of the
assumptions in some ways.

One also has to recall that gravity has no local gauge-invariant degrees of
freedom. So one cannot start with a theory containing local gauge-invariant
operators to generate Einstein gravity as an emergent phenomenon
in the {\it same} spacetime.

The clue to relax the first assumption comes from the holographic
principle \cite{holography}:
{\it Gravitons live in a higher dimensional spacetime than gauge fields}.
In this description known as the anti-de Sitter space and conformal field
theory (AdS/CFT) duality \cite{ads-cft},
gravity in higher dimensions is an emergent phenomenon arising
from particle interactions in a gravityless, lower-dimensional
world. The bulk geometry, i.e., AdS space, has to contain a holographic
screen on which the boundary theory, typically super Yang-Mills theory,
is defined. Since the Yang-Mills theory contains many
local gauge-invariant observables and the diffeomorphism only
appears in the bulk, the AdS/CFT duality does not provide any
information on lower-dimensional gravity in the same spacetime
where the boundary theory is defined.

One may try to relax the second assumption instead.
This assumption can be relaxed, for example, by introducing a symplectic
structure of spacetime -- noncommutative (NC) spacetime:
\be \la{nc-space}
[y^\mu, y^\nu]_\star = i \theta^{\mu\nu}.
\ee
A field theory defined on the NC spacetime preserves neither locality
nor Lorentz invariance. An important fact is that translations in NC directions
are basically gauge transformations, i.e.,
\begin{equation} \label{inner}
e^{ik \cdot y} \star f(y) \star e^{-ik \cdot y} = f(y + \theta \cdot k).
\end{equation}
This immediately implies that there are no local gauge-invariant
observables in NC gauge theory \cite{gross}. These properties are precisely
those of gravity. This was the motivation in \ct{rivelles} to explore the
relation between NC field theory and gravity.

Recently it was shown in \ct{sty,ys} that gravitational instantons in Einstein gravity
are equivalent to $U(1)$ instantons in NC gauge theory, so illuminating
the remarkable correspondence between NC field theory and gravity at least
for their self-dual sectors.

In this paper we will generalize the results in \ct{sty,ys}, showing
that self-dual electromagnetism in NC spacetime is equivalent to
self-dual Einstein gravity. Although the equivalence between NC
instantons and gravitational instantons was formally shown in
\ct{sty,ys}, an explicit example was shown only for the
Eguchi-Hanson metric and a single NC instanton. Furthermore it is a
highly nontrivial problem to find multi-instanton solutions even in
the gauge theory. Therefore one might expect that it will be very
difficult to explicitly find a map that sends a generic self-dual
gauge field configuration of NC gauge theory to a metric which is a
solution of self-dual Einstein equations. It is thus quite
surprising that there is an elegant formulation to realize the
equivalence between self-dual NC electromagnetism and self-dual
Einstein gravity, where it is not difficult to find explicit
nontrivial solutions showing that the asymptotically locally
Euclidean (ALE) and asymptotically locally flat (ALF) spaces
\ct{gibb-hawk} as well as the real heaven \ct{real-heaven} are indeed solutions of NC electromagnetism.

This formulation is essentially based on Eq.\eq{inner} stating that translations are
an inner automorphism of NC algebra \eq{nc-space} or, in its infinitesimal form,
derivations are inner, i.e.,
\be \la{inner-derivation}
-i [(\theta^{-1})_{\mu\nu} y^\nu, f(y)]_\star = \partial_\mu f(y).
\ee
In the presence of gauge fields, the coordinates $y^\mu$
should be promoted to the covariant coordinates defined by
\begin{equation}\label{cov-coord}
 x^\mu(y) \equiv y^\mu + \theta^{\mu\nu} \widehat{A}_\nu (y)
\end{equation}
in order for star multiplications to preserve the gauge covariance \ct{madore}.
The inner derivations \eq{inner-derivation} are accordingly covariantized
too as follows
\bea \la{vector-map}
{\rm ad}_{D_\mu}[f] &\equiv& -i [D_\mu(y), f(y)]_\star = \theta^{ab}
\frac{\partial D_\mu}{\partial y^a}\frac{\partial f}{\partial y^b} + \cdots \xx
&\equiv& V_\mu^a (y) \partial_a f(y) + {\cal O}(\theta^3)
\eea
where $D_\mu (y) \equiv (\theta^{-1})_{\mu\nu} x^\nu (y)$. It turns
out that the vector fields $V_\mu (y) \equiv V_\mu^a(y)
\partial_a$ form an orthogonal frame and hence define vielbeins of a
gravitational metric. This intrinsic property in NC geometry will be
combined with the rigorously established result in
\ct{ashtekar,mason-newman,beautiful} to conclude that self-dual NC
electromagnetism is equivalent to self-dual Einstein gravity, which
will be done in Section 3.

Self-dual gauge fields in NC spacetime have been shown in
\ct{sty,ys} to describe hyper-K\"ahler manifolds and every hyper-K\"ahler manifold
has an associated twistor space \ct{hklr}. Therefore we expect that
self-dual NC electromagnetism may be related to a twistor space
describing curved self-dual spacetime. We will show that this
structure is beautifully realized through the Darboux theorem, where
all the information of self-dual NC gauge fields can be encoded in
the holomorphic structure of the twistor space. This construction
also clarifies the nature of gravity emerging from NC gauge fields;
the gauge fields act as a deformation of the K\"ahler structure of
self-dual 4-manifold or the complex structure of twistor space. In
this way gauge fields in NC spacetime manifest themselves as a
deformation of background geometry, which is also compatible with
the picture in Eq.\eq{vector-map}. Thus we should think of the
twistor space as already incorporating the backreaction of NC
instantons. This picture is very similar to that in
\ct{twistor-string} where placing D1-branes (as instantons in gauge theory)
in twistor space is interpreted as blowing up points in four dimensions
dubbed as spacetime foams and the K\"ahler blowups in four dimensions
are encoded in the twistor space as the backreaction of the D1-branes.

A brief outline of the paper is the following. In Section 2, the
equivalence between NC $U(1)$ instantons and gravitational
instantons is made more precise giving a nice geometrical picture in
terms of the twistor geometry \ct{penrose}. The beautiful
geometrical structures of twistor geometry also allow us to clarify
the nature of gravity emerging from NC gauge fields. In Section 3,
we apply Eq.\eq{vector-map} to self-dual NC gauge fields to show the
equivalence between self-dual electromagnetism in NC spacetime and
self-dual Einstein gravity, using the rigorously established result
in \ct{ashtekar,mason-newman,beautiful}. In Section 4, we add some
remarks about our results.

\section{2. Noncommutative instantons and twistor space}

Let us consider electromagnetism in the NC spacetime defined by Eq.\eq{nc-space}.
The action for the NC $U(1)$ gauge theory in flat Euclidean ${\bf R}^4$
is given by
\begin{equation}\label{nced}
\widehat{S}_{\mathrm{NC}} = \frac{1}{4}\int\! d^4 y \,\widehat{F}_{\mu\nu}
\star \widehat{F}^{\mu\nu},
\end{equation}
where NC electromagnetic fields are defined by
\be \la{ncf}
\widehat{F}_{\mu\nu}=\partial_{\mu}\widehat{A}_{\nu}-\partial_{\nu}\widehat{A}_{\mu}-
i \,[\widehat{A}_{\mu}, \widehat{A}_{\nu}]_\star.
\ee
Contrary to ordinary electromagnetism, the NC $U(1)$ gauge theory admits
non-singular instanton solutions satisfying
the NC self-duality equation \ct{nek-sch},
\be \la{nc-self-dual}
{\widehat F}_{\mu\nu} (y) = \pm \half
\varepsilon_{\mu\nu\lambda\sigma}{\widehat F}_{\lambda\sigma} (y).
\ee

The NC gauge theory has an equivalent dual description through the so-called Seiberg-Witten
(SW) map in terms of ordinary gauge theory on commutative spacetime \ct{sw}.
The SW map is a map between gauge orbit spaces of commutative and NC gauge
fields. We will mainly be confined to semi-classical limit, say ${\cal O}(\theta)$,
which means slowly varying fields, $\sqrt{\kappa} |\frac{\partial F}{F}| \ll 1$,
in the sense keeping field strengths (without restriction on their
size) but not their derivatives. This is precisely the
limit taking only the leading term in Eq.\eq{vector-map},
so ignoring derivative corrections which start from ${\cal O}(\theta^3)$.
(See \ct{paper} for derivative corrections of the SW map.)
The exact SW map \ct{sw-map,ban-yang} in this limit is given by
\bea \label{eswmap}
    &&  \widehat{F}_{\mu\nu}(y) = \Bigl(\frac{1}{1 + F\theta} F
    \Bigr)_{\mu\nu}(x), \\
    \la{measure-sw}
    && d^{4} y = d^{4} x \sqrt{\det(1+ F \theta)}(x),
\eea
where $x^\mu$'s are the covariant coordinates defined by Eq.\eq{cov-coord}.
Applying the maps \eq{eswmap} and \eq{measure-sw} to the action \eq{nced},
one can get the commutative nonlinear electrodynamics \ct{ban-yang,yang}
equivalent to Eq.\eq{nced}
\begin{equation}\label{ced-sw}
S_{\mathrm{C}} = \frac{1}{4} \int d^4 x \sqrt{\det{{\rm g}}} \;
{\rm g}^{\mu \lambda} {\rm g}^{\sigma\nu} F_{\mu\nu}
F_{\lambda\sigma},
\end{equation}
where we introduced an ``effective metric" induced
by the dynamical gauge fields as
\begin{equation}\label{effective-metric}
    {\rm g}_{\mu\nu} = \delta_{\mu\nu} + (F\theta)_{\mu\nu},
    \;\;  ({\rm g}^{-1})^{\mu\nu} \equiv {\rm g}^{\mu\nu} =
    \Bigl(\frac{1}{1 + F\theta}\Bigr)^{\mu\nu}.
\end{equation}

It was shown in \ct{sty} that the self-duality equation for the action
$S_{\mathrm{C}}$ is given by
\be \la{c-self-dual}
{\bf F}_{\mu\nu} (x) = \pm \half
\varepsilon_{\mu\nu\lambda\sigma}{\bf F}_{\lambda\sigma} (x),
\ee
where
\begin{equation}\label{def-fatf}
{\bf F}_{\mu\nu} (x) \equiv \Bigl({\rm g}^{-1} F \Bigr)_{\mu\nu} (x).
\end{equation}
Note that Eq.\eq{c-self-dual} is nothing but the exact SW map \eq{eswmap}
of the NC self-duality equation \eq{nc-self-dual}.
It turned out \ct{sty,ys} that Eq.\eq{c-self-dual} describes gravitational
instantons obeying the self-dual equations \ct{g-instanton}
\be \la{instanton-gravity}
R_{abcd} = \pm \half \varepsilon_{abef}
{R^{ef}}_{cd},
\ee
where $R_{abcd}$ is a curvature tensor. Interestingly, Eq.\eq{c-self-dual} can
be rewritten \ct{paper} as the self-duality in a curved space described
by the metric ${\rm g}_{\mu\nu}$
\be \la{curved-self-dual}
F_{\mu\nu} (x) = \pm \half \frac{\varepsilon^{\lambda\sigma\rho\tau}}
{\sqrt{\det{{\rm g}}}} {\rm g}_{\mu\lambda}{\rm g}_{\nu\sigma}
F_{\rho\tau} (x).
\ee
It should be remarked, however, that the self-duality in \eq{curved-self-dual}
cannot be interpreted as a usual self-duality equation
in a fixed background since the four-dimensional metric used to define
Eq.\eq{curved-self-dual} depends in turn on the $U(1)$ gauge fields.

NC instantons turn out to have a rich connection with twistor geometry
\ct{penrose}. We will newly derive the key results in \ct{sty,ys}
closely following the paper \ct{ooguri-vafa} on $N=2$ strings which
will clarify the nature of gravity emerging from NC gauge fields. To
proceed in that direction, let us consider the line element (where
we follow the notations in \ct{sty,ys})
\be \la{dia-metric}
ds^2 = \widetilde{{\rm g}}_{\mu\nu} dx^\mu dx^\nu =
 \widetilde{\sigma}_\mu \otimes  \widetilde{\sigma}_\mu
\ee
where ${\rm g}_{\mu\nu}= 1/2(\delta_{\mu\nu} + \widetilde{{\rm g}}_{\mu\nu})$.
We also introduce the triple of K\"ahler forms as follows,
\be \la{unit-2-form}
\widetilde{\omega}^a = \half \eta^a_{\mu\nu} \widetilde{\sigma}^\mu
\wedge \widetilde{\sigma}^\nu.
\ee
It is easy to check \ct{ys} that $\widetilde{\sigma}_1 \wedge  \widetilde{\sigma}_2 \wedge
\widetilde{\sigma}_3 \wedge  \widetilde{\sigma}_4 = d^4 x$, say,
$\sqrt{\det{\widetilde{{\rm g}}_{\mu\nu}}} = 1$ and
\bea \la{triple-kahler}
&& \omega \equiv \widetilde{\omega}^2 + i \widetilde{\omega}^1 = dz_1 \wedge dz_2,
\; \bar{\omega} \equiv \widetilde{\omega}^2 - i \widetilde{\omega}^1
=  d\zbar_1 \wedge d\zbar_2, \xx
&& \Omega \equiv - \widetilde{\omega}^3 = \frac{i}{2}(dz_1 \wedge d\zbar_1
+ dz_2 \wedge d\zbar_2) + \theta F.
\eea
It is obvious that $d\widetilde{\omega}^a = 0, \; \forall a$ because of the
Bianchi identity, $dF=0$. This means that
the metric $\widetilde{{\rm g}}_{\mu\nu}$ is hyper-K\"ahler \ct{ys}, which is
an equivalent statement as Ricci-flat K\"ahler in four dimensions.
Therefore the metric $\widetilde{{\rm g}}_{\mu\nu}$ is a gravitational
instanton \ct{g-instanton}.
Eq.\eq{triple-kahler} clearly shows how dynamical gauge fields living in NC
spacetime deform the K\"ahler structure and thus induce a deformation of
background geometry through gravitational instantons,
thus realizing the emergent geometry.

The deformation of symplectic (or K\"ahler) structure on ${\bf R}^4$ due to
the fluctuation of gauge fields can be more clarified by the following construction.
Let us consider a deformation of the holomorphic (2,0)-form
$\omega = dz_1 \wedge dz_2$ as follows
\be \la{deformation}
\Psi (t) = \omega + i t \Omega + \frac{t^2}{4} \bar{\omega}
\ee
where the parameter $t$ takes values in ${\bf P}^1={\bf S}^2$.
One can easily see that $d\Psi(t) = 0$ if and only if $dF=0$ and
\be \la{psi-psi}
\Psi (t) \wedge \Psi (t) = 0
\ee
since Eq.\eq{psi-psi} is equivalent to the instanton equation
$F_{\mu\nu}^+ = 1/4(F\widetilde{F}) \theta_{\mu\nu}^+$ \ct{sty,ys}.
Since the two-form $\Psi(t)$ is closed and degenerate, the Darboux theorem
asserts that one can find a $t$-dependent map $(z_1,z_2)  \to (Z_1(t;z_i,
\zbar_i), Z_2(t;z_i, \zbar_i))$ such that
\be \la{symplectic-form}
\Psi (t) = dZ_1(t;z_i,\zbar_i) \wedge dZ_2(t;z_i, \zbar_i).
\ee

When $t$ is small, one can solve \eq{symplectic-form} by expanding
$Z_i(t;z,\zbar)$ in powers of $t$ as
\be \la{small-expansion}
Z_i(t;z,\zbar) = z_i + \sum_{n=1}^{\infty} \frac{t^n}{n}p_n^i(z,\zbar).
\ee
By substituting this into Eq.\eq{deformation}, one gets at ${\cal O}(t)$
\bea \la{exp-eq1}
&& \partial_{z_i} p_1^i = 0, \\
\la{exp-eq2}
&& \epsilon_{ik} \partial_{\zbar_j} p_1^k dz^i \wedge d\zbar^j= i \Omega
\eea
where the fact was used that $\Omega$ is a (1,1)-form.
Eq.\eq{exp-eq1} can be solved by setting $p_1^i = 1/2 \epsilon^{ij} \partial_{z_j}
K$ and then $\Omega = i/2  \partial_i {\bar \partial_j} K dz^i \wedge
d\zbar^j$. The real-valued smooth function $K$ is the K\"ahler
potential of $U(1)$ instantons in \ct{sty,ys}. In terms of this K\"ahler two-form
$\Omega$, Eq.\eq{psi-psi} reduces to the complex Monge-Amp\`ere or
the Pleba\'nski equation \ct{plebanski}
\be \la{cma-pleb}
\Omega \wedge \Omega = \half \omega \wedge \bar{\omega},
\ee
that is, $\det(\partial_i {\bar \partial_j} K) =1 $.

When $t$ is large, one can introduce another Darboux coordinates
${\widetilde Z}_i(t;z_i,\zbar_i)$ such that
\be \la{symplectic-form2}
\Psi (t) = t^2 d{\widetilde Z}_1(t;z_i,\zbar_i) \wedge
d{\widetilde Z}_2(t;z_i, \zbar_i)
\ee
with expansion
\be \la{small-expansion2}
{\widetilde Z}_i(t;z,\zbar) = \zbar_i + \sum_{n=1}^{\infty} \frac{t^{-n}}{n}
{\widetilde p}_n^i(z,\zbar).
\ee
One can get the solution \eq{deformation} with ${\widetilde p}_1^i = - 1/2
\epsilon^{ij} \partial_{\zbar_j} K$ and $\Omega = i/2  \partial_i {\bar \partial_j} K dz^i \wedge
d\zbar^j$.

The $t$-dependent Darboux coordinates $Z_i(t;z,\zbar)$ and
${\widetilde Z}_i(t;z,\zbar)$ correspond to holomorphic coordinates
on two local charts, where the 2-form $\Psi(t)$ becomes the holomorphic
(2,0)-form, of the dual projective twistor space $P^* T$
which may be viewed as a fiber bundle over ${\bf S}^2$
with a fiber being a hyper-K\"ahler manifold ${\cal M}_4$.
Here we regard $t$ as a deformation parameter of complex structure
on ${\cal M}_4$. We know from Eq.\eq{triple-kahler} that
the K\"ahler form $\Omega$ is rank 4 and so can always serve as a symplectic
form on both coordinate charts. As a result, two sets of coordinates at $t=0$ and
$t=\infty$ are related to each other by a $t$-dependent
canonical transformation given by $f_i(t;Z(t))=t {\widetilde Z}_i(t)$
on an overlapping coordinate chart \ct{beautiful,ooguri-vafa}.
In this way, the complex geometry of the twistor space $P^*T$ encodes all
the information about the K\"ahler geometry of self-dual 4-manifolds ${\cal
  M}_4$ emergent from self-dual NC gauge fields.
We refer the reader to \ct{penrose} for more details about the
twistor geometry.

From the above construction, we see that the information of self-dual NC gauge
fields is encoded in the twistor space in the following way.
The Bianchi identity $dF=0$ appears as the K\"ahler condition of a self-dual
metric on ${\cal M}_4$ or as $d\Psi(t)=0$ in the twistor space.
The self-duality equation \eq{c-self-dual} is realized
as the Ricci flatness of ${\cal M}_4$ or as the nilpotentness \eq{psi-psi}
in the twistor space. In total, the self-dual NC gauge fields manifest
themselves as a Ricci flat K\"ahler manifold ${\cal M}_4$ or
as a holomorphic deformation of the twistor space.
If turning off gauge fields, namely $F=0$, we simply arrive at a flat space,
i.e., ${\cal M}_4 = {\bf R}^4$ and $P^*T= {\bf R}^4 \times {\bf S^2}$.
Thus we should think of the twistor space as already incorporating the backreaction of
NC instantons. This picture is remarkably similar to that in twistor string
theory \ct{twistor-string} as we discussed in Section 1.

\section{3. Self-dual Einstein gravity from NC electromagnetism}

In this section we will use the background independent formulation of NC gauge
theories \ct{sw,seiberg}. One can show using the SW map in \ct{ban-yang,yang}
for this case that
\bea \label{sw-gen}
&& \int d^4 y (\widehat{F}-B)_{\mu\nu}
\star   (\widehat{F}-B)^{\mu\nu} \xx
&=& \int d^4 x \sqrt{\det{\mathrm{g}}} \;
{\mathrm{g}}^{\mu \lambda} {\mathrm{g}}^{\sigma\nu} B_{\mu\nu} B_{\lambda\sigma}
\eea
where $B_{\mu\nu} = (\theta^{-1})_{\mu\nu}$. The identity
(\ref{sw-gen}) definitely shows that the fluctuations $\widehat{F}$
by NC photons around the background $B$ are mapped through the SW
map to the fluctuations of geometry on commutative spacetime.

Before going into details, let us point out that the NC gauge theory
and gravity correspondence may be understood as a large $N$ duality
\ct{mine}. To see this picture, consider the NC description in
\ct{sw,seiberg} where the action is expressed only in terms of
manifestly covariant and background-independent variables:
\begin{eqnarray} \label{matrix-sw}
&& \frac{1}{4G_s}\int d^4 y (\widehat{F}-B)_{\mu\nu}
\star   (\widehat{F}-B)^{\mu\nu} \xx
&=& -\frac{\pi^2}{g_s \kappa^2} g_{\mu\lambda} g_{\nu\sigma}
{\mathbf{Tr}}_{\mathcal{H}} [x^\mu,x^\nu][x^\lambda, x^\sigma]
\end{eqnarray}
where we made a replacement $\frac{1}{(2\pi)^2} \int
\frac{d^4y}{|\mathrm{Pf}\theta|} \leftrightarrow  \mathbf{Tr}_{\mathcal{H}}$ using the Weyl-Moyal map.
Here we used the following notation: $g_{\mu\nu}$ = constant closed
  string metric, $g_s (G_s)$ = closed (open) string coupling constant,
$\kappa = 2 \pi \alpha^\prime$.
The covariant, background-independent coordinates $x^\mu$ \ct{madore,seiberg}
are defined by Eq.\eq{cov-coord} and they are now operators
acting on an infinite-dimensional, separable Hilbert
space $\mathcal{H}$, which is the representation space of the Heisenberg
algebra \eq{nc-space}. The NC gauge symmetry in Eq.(\ref{matrix-sw}) then acts
as unitary transformations on $\mathcal{H}$, i.e.,
\be \la{nc-symmetry}
x^\mu \rightarrow {x^{\prime}}^\mu = U x^\mu U^\dagger.
\ee
This NC gauge symmetry $U_{\rm{cpt}}(\mathcal{H})$ is so large that
$U_{\rm{cpt}}({\mathcal{H}}) \supset U(N) \;(N \rightarrow \infty)$
\cite{harvey}. In this sense the NC gauge theory in
Eq.(\ref{matrix-sw}) is indeed a large $N$ gauge theory. Note that
the second expression in Eq.\eq{matrix-sw} is a large $N$ version of
the IKKT matrix model which describes the nonperturbative dynamics
of type IIB string theory \ct{ikkt}.

The map \eq{vector-map} in general defines an inner derivation of NC
$\star$-algebra which reduces to ordinary vector fields or Lie
derivatives in the limit of slowly varying fields. Therefore we will
identify the adjoint action of $D_\mu$ with respect to the
star-product with vector fields $V_\mu \in T{\cal M}_4$ on some four
manifold ${\cal M}_4$ as an emergent geometry. The identification is
quite natural since the NC gauge fields $\widehat{A}_\mu(y)$ are in
general arbitrary, so they generate arbitrary vector fields in
$T{\cal M}_4$ according to the map \eq{vector-map} and $-i
[B_{\mu\nu} x^\nu, f]_\star = \partial_\mu f$ when $\widehat{A}_\mu
= 0$. One can easily check that
\bea \la{lie-bracket}
&& ({\rm ad}_{D_\mu}\, {\rm ad}_{D_\nu}-{\rm ad}_{D_\nu}\,{\rm
ad}_{D_\mu})[f] \xx &=& - i \, {\rm ad}_{[D_\mu, D_\nu]_\star}[f]
\approx [V_\mu,V_\nu][f]
\eea
where the right-hand side is defined by the Lie bracket between
vector fields in $T{\cal M}_4$.

Now let us look for an instanton solution of Eq.(\ref{matrix-sw}).
Since the instanton is a Euclidean solution with a finite action,
the instanton configuration should approach to a pure gauge at
infinity. Our boundary condition is $\widehat{F}_{\mu\nu} \to 0$ at
$|y| \to \infty$ as usual. Thus one has to remove the background
part from the action \eq{matrix-sw}, but with a background
independent as well as gauge covariant way. There is a unique way to
achieve this by defining the self-duality equation as follows
\bea \la{sde-matrix}
&& {\rm ad}_{[D_\mu, D_\nu]_\star} =
\pm \half \varepsilon_{\mu\nu\lambda\sigma} \; {\rm ad}_{[D_\lambda, D_\sigma]_\star}
\Leftrightarrow \xx
&& [V_\mu,V_\nu] =
\pm \half \varepsilon_{\mu\nu\lambda\sigma} [V_\lambda, V_\sigma],
\eea
where we used Eq.\eq{lie-bracket}. From the above definition, it is
obvious that the constant part in $ -i [D_\mu, D_\nu]_\star =
{\widehat F}_{\mu\nu} - B_{\mu\nu}$, i.e. $B_{\mu\nu}$, can be
dropped and so the self-duality equation \eq{sde-matrix} is
equivalent to Eq.\eq{nc-self-dual}.

It is obvious from Eq.\eq{vector-map} that the vector fields $V_\mu$
are divergence free, i.e. $\partial_a V^a_\mu = 0$. In other words,
they preserve a fixed volume form $\vare_4$, i.e., ${\cal L}_{V_\mu}
\vare_4 = 0, \;\forall \mu$, where ${\cal L}_{V_\mu}$ is the Lie derivative along
$V_\mu$. In addition, the vector fields $V_\mu$ form an orthonormal
frame up to some conformal factor since $\vare_4(V_1,V_2,V_3,V_4)
\propto \det (\frac{\partial x}{\partial y} )$ corresponds to the Jacobian
factor for the coordinate transformation, $y^\mu \to x^\mu(y)$, which is
non-vanishing unless $F+B=0$.
In summary, instanton configurations in NC spacetime \eq{nc-space}
are mapped to the volume preserving diffeomorphism,
denoted as $SDiff({\cal M}_4)$, satisfying Eq.\eq{sde-matrix}.

So we arrive at the result of Ashtekar {\it et al.} \cite{ashtekar}.
Their result is summarized as follows \ct{joyce}.
Let ${\cal M}_4$ be an oriented 4-manifold and let $V_\mu$ be
vector fields on ${\cal M}_4$ forming an oriented basis for $T{\cal M}_4$.
Then $V_\mu$ define a conformal structure $[{\cal G}]$ on ${\cal M}_4$.
Suppose that $V_\mu$ preserve a volume form on ${\cal M}_4$
and satisfy the self-duality equation
\begin{equation} \label{ashtekar}
[ V_\mu, V_\nu ] = \pm \frac{1}{2} {\varepsilon}_{\mu\nu\lambda\sigma}
[ V_\lambda, V_\sigma ].
\end{equation}
Then $[{\cal G}]$ defines an (anti-)self-dual and Ricci-flat metric.

Since $V_\mu \in LSDiff({\cal M}_4)$, the Lie algebra of
$SDiff({\cal M}_4)$, form an orthonormal frame up to a conformal
factor, they define a metric on ${\cal M}_4$. Indeed $V_\mu^a$ in
Eq.\eq{vector-map} can be related to an inverse vierbein or tetrad
on ${\cal M}_4$. Thus the self-dual metric is given by
\ct{ashtekar}
\be \la{sd-metric}
{\cal G}^{ab} = {\det V}^{-1}V_\mu^a V_\nu^b
    \delta^{\mu\nu}.
\ee
Note that the metric \eq{sd-metric} becomes flat if NC gauge fields
are trivial, say, $\widehat{A}_\mu = 0$ since $V_\mu^a =
\delta_\mu^a$ in this case. In other words, the flat spacetime is
emergent from the uniform condensation of gauge fields, i.e.,
$B_{\mu\nu} = (1/\theta)_{\mu\nu}$, which defines the NC spacetime
\eq{nc-space} as a vacuum. But Eq.\eq{sd-metric} implies
that the presence of nontrivial gauge fields in the NC spacetime
deforms the vacuum manifold, i.e. flat ${\bf R}^4$, to a curved
manifold. We here confirm again the same picture emergent in Section
2 that NC instantons deform ${\bf R}^4$ to a self-dual Einstein
manifold.

Motivated by the similarity of Eq.\eq{ashtekar} to the self-duality
equation of Yang-Mills theory, Mason and Newman showed
\cite{mason-newman} that, if we have a reduced Yang-Mills theory
where the gauge fields take values in the Lie algebra of
$SDiff({\cal M}_4)$, which is exactly the case in
Eq.\eq{sde-matrix}, Yang-Mills instantons are actually equivalent to
gravitational instantons. We just showed that this is the case for
NC electromagnetism.

It is {\it a priori} not obvious that the self-dual electromagnetism
in NC spacetime is equivalent to the self-dual Einstein gravity.
Therefore it should be helpful to have explicit nontrivial examples
to appreciate how the equivalence is achieved. It is not difficult
to find explicit self-dual solutions from Eq.\eq{ashtekar}. For
example, the Gibbons-Hawking metric \ct{gibb-hawk} was given in
\ct{joyce} and the real heaven solution \ct{real-heaven} was done in
\ct{real-heaven2}.

The Gibbons-Hawking metric \ct{gibb-hawk} is a general class of
self-dual, Ricci-flat metrics with the triholomorphic $U(1)$
symmetry which describes a particular class of ALE and ALF
instantons. Let $(a_i, U),\; i=1,2,3,$ are smooth real functions on
${\bf R}^3$ and define $V_i = - a_i \frac{\partial}{\partial \tau} +
\frac{\partial}{\partial x^i}$ and $V_4 = U \frac{\partial}{\partial
\tau}$, where $\tau $ parameterizes a circle and $\vec{x} \in {\bf
R}^3$. Then one can easily check that $(V_i,V_4)$ are
divergence-free, namely, they belong to $LSDiff({\cal M}_4)$ and the
Killing vector $\partial / \partial \tau$ generates the
triholomorphic $U(1)$ symmetry. Eq.(\ref{ashtekar}) then reduces to
the equation $ \nabla U +
\nabla \times \vec{a} = 0$ and the metric whose inverse
is Eq.\eq{sd-metric} is given by
\be \label{gh-metric}
ds^2 = U^{-1}(d\tau + \vec{a} \cdot d\vec{x})^2 + U d\vec{x} \cdot d\vec{x}.
\ee

The real heaven metric \ct{real-heaven} describes four dimensional
hyper-K\"ahler manifolds with a rotational Killing symmetry which is
also completely determined by one real scalar field.
The vector fields $V_\mu$ in this case are given by \ct{real-heaven2}
\bea \la{heaven-vector}
&& V_1 = \frac{\partial}{\partial x^1} - \partial_2 \psi \frac{\partial}{\partial \tau} \xx
&& V_2 = \frac{\partial}{\partial x^2} + \partial_1 \psi \frac{\partial}{\partial \tau} \\
&& V_3 = e^{\psi/2} \left ( \sin \Bigl(\frac{\tau}{2} \Bigr)
\frac{\partial}{\partial x^3} + \partial_3 \psi \cos
\Bigl(\frac{\tau}{2} \Bigr) \frac{\partial}{\partial \tau}  \right)
\xx && V_4 = e^{\psi/2} \left( \cos \Bigl(\frac{\tau}{2}\Bigr)
\frac{\partial}{\partial x^3} - \partial_3 \psi \sin
\Bigl(\frac{\tau}{2} \Bigr) \frac{\partial}{\partial \tau} \right)
\nonumber
\eea
where the rotational Killing vector is given by $c_i \partial_i \psi \partial/\partial \tau$
with constants $c_i \; (i=1,2)$ and the function $\psi$
is independent of $\tau$. Then it is easy to see that $V_\mu \in LSDiff({\cal M}_4)$.
Eq.(\ref{ashtekar}) is now equivalent to
the three-dimensional continual Toda equation $(\partial_1^2 + \partial_2^2) \psi
+ \partial_3^2 e^\psi = 0$ and the metric is determined by Eq.\eq{sd-metric} as
\be \label{heaven-metric}
ds^2 = (\partial_3 \psi)^{-1}(d\tau +  a^i d x^i)^2 + (\partial_3 \psi)
( e^\psi dx^i dx^i + dx^3 dx^3)
\ee
where $a^i = \varepsilon^{ij} \partial_j \psi$.

Although we concretely derived the equivalence, Eq. \eq{sde-matrix},
between self-dual NC electromagnetism and self-dual Einstein gravity,
it is nevertheless very surprising that the real heaven \eq{heaven-metric} as well as
the ALE and ALF instantons \eq{gh-metric} are solutions of NC electromagnetism.

\section{4. Discussion}

We have shown a remarkable picture implying that gravity can emerge
from gauge fields living in NC spacetime. This striking picture
about gravity is originated from the fact that the fluctuation of
gauge fields in NC spacetime can be interpreted as that of spacetime
geometry via the Darboux theorem. It turns out
\ct{mine,paper} that a basic reason for the correspondence between
NC gauge theory and gravity is that the NC spacetime \eq{nc-space}
admits an extra symmetry, the so-called $\Lambda$-symmetry (or
$B$-field transformation), which is as large as diffeomorphism
symmetry. Through the Darboux theorem, the $\Lambda$-symmetry can be
considered as a par with diffeomorphisms. This is the underlying
reason why gravity can emerge from gauge fields living in NC
spacetime. One can also see this picture through the SW map in
Eq.\eq{sw-gen}.

Some remarks are in order.

(1) Recently, it was found \ct{twist-symm} that NC field theory is
invariant under the twisted Poincar\'e symmetry where the action of
generators is now defined by the twisted coproduct in the deformed
Hopf algebras. Especially the NC spacetime \eq{nc-space} appears as
a twisted Poincar\'e invariant as well as gauge invariant (since
$\theta^{\mu\nu}=(1/B)^{\mu\nu}$) while it has been argued to break
the usual Lorentz symmetry. This fact seems to be consistent with
the picture of emergent gravity, as we observed below
Eq.\eq{sd-metric}, where a flat spacetime is not {\it a priori}
given but emergent from the algebra \eq{nc-space} defining a NC
spacetime. According to the motto of emergent gravity declaring that
a geometry is defined by an algebra, we interpret that the NC
algebra \eq{nc-space} defines a flat spacetime as a vacuum geometry
(i.e., without fluctuating fields) in the context of emergent
geometry. If it is true, it implies that a flat spacetime as well as
its Lorentz symmetry is emergent from the NC $\star$-algebra
\eq{nc-space}, which is twisted Poincar\'e invariant as well as
gauge invariant. (See the Seiberg's paper in \ct{review} for a
related argument.) We think that the twisted Poincar\'e symmetry,
especially the deformed Hopf algebra and quantum group structures,
will be important to understand the correspondence between NC field
theory and gravity since underlying symmetries are always an
essential guide for physics. Incidentally, this symmetry plays a
prominent role to construct NC gravity \ct{nc-gravity}.

(2) The canonical structures, in particular, complex and K\"ahler
structures, of the self-dual system \eq{ashtekar}, have been fully
studied in a beautiful paper \ct{beautiful}. The arguments in
\ct{beautiful} are essentially the same as ours leading to
Eq.\eq{cma-pleb}. It was also shown there how Pleba\'nski's heavenly
equations \ct{plebanski} can be derived from Eq.\eq{ashtekar} and
also how Eq.\eq{ashtekar} manifests the structure of twistor space.
Furthermore it can be shown \ct{husain} that Eq.\eq{ashtekar} can be
reduced to the $sdiff(\Sigma_g)$ chiral field equations in two
dimensions, where $sdiff(\Sigma_g)$ is the area preserving
diffeomorphisms of a Riemann surface of genus $g$. All these
properties are deeply related to several integrable structures. Thus
it will be very interesting to clarify the integrable structures
inherent in the self-dual NC electromagnetism, self-dual gravity and
twistor spaces.

\section*{Acknowledgements}

We would like to thank Chong-Sun Chu, Harald Dorn, Giuseppe Policastro,
Mario Salizzoni and Alessandro Torrielli for helpful discussions.
This work was supported by the Alexander von Humboldt Foundation.

\end{document}